\documentclass[aps,showpacs,prb,reprint,groupaddress,longbibliography]{revtex4-1}

\usepackage[colorlinks=true,linkcolor=blue,citecolor=blue,urlcolor=blue]{hyperref}

\usepackage{hyperref}
\usepackage{setspace} 
\usepackage{graphicx}
\usepackage{amsmath}
\usepackage{color}
\usepackage{amsmath}
\usepackage{amssymb}
\usepackage{verbatim}
\usepackage{latexsym}
\usepackage{enumerate} 
\usepackage{bm} 

\newcommand{\argmin}{\operatornamewithlimits{arg\,min}}

\begin{document}

\title{Correlation effects on electron-phonon coupling in semiconductors: many-body theory along thermal lines} 

\author{Bartomeu Monserrat}
\email{bm418@cam.ac.uk}
\affiliation{Department of Physics and Astronomy, Rutgers University,
  Piscataway, New Jersey 08854-8019, USA}
\affiliation{TCM Group, Cavendish Laboratory, University of Cambridge,
  J.\ J.\ Thomson Avenue, Cambridge CB3 0HE, United Kingdom}

\date{\today}

\begin{abstract}
A method is proposed for the inclusion of electron correlation in the calculation of the temperature dependence of band structures arising from electron-phonon coupling. It relies on an efficient exploration of the vibrational phase space along the recently introduced thermal lines. Using the $G_0W_0$ approximation, the temperature dependence of the direct gaps of diamond, silicon, lithium fluoride, magnesium oxide, and titanium dioxide is calculated. Within the proposed formalism, a single calculation at each temperature of interest is sufficient to obtain results of the same accuracy as in alternative, more expensive methods. It is shown that many-body contributions beyond semi-local density functional theory modify the electron-phonon coupling strength by almost $50$\% in diamond, silicon, and titanium dioxide, but by less than $5$\% in lithium flouride and magnesium oxide. The results reveal a complex picture regarding the validity of semi-local functionals for the description of electron-phonon coupling.  
\end{abstract}


\maketitle



The coupling of electrons and atomic vibrations in solids underlies many phenomena, such as phonon-mediated superconductivity~\cite{cooper_elph,BCS_theory}, the temperature dependence of band structures~\cite{RevModPhys.77.1173}, electronic transport~\cite{gw_graphene_transport,verstraete_transport,walle_transport}, or phonon-assisted optical absorption~\cite{ph_assited_optical,ph_assited_optical_noffsinger,ph_assisted_optical_zacharias}. Almost all first principles calculations of electron-phonon coupling (EPC) rely on semi-local density functional theory (DFT)~\cite{PhysRevLett.105.265501,gonze_off_diagonal,PhysRevB.87.144302,elph_Si_nano,gonze_marini_elph,monserrat_elph_diamond_silicon,patrick_molecule_solid_2014,giustino_nat_comm,helium,PhysRevLett.107.255501,molec_crystals_elph}. However, recent studies have shown that many-body effects are required to accurately capture the EPC strength in carbon-based materials~~\cite{graphene_epc_beyond_dft_aleiner,graphene_epc_beyond_dft_mauri,graphite_epc_beyond_dft_mauri,c60_epc_beyond_dft,gw_graphene_transport,gonze_gw_elph}, and in some families of superconductors~\cite{kotliar_epc_beyond_dft,haule_epc_dmft,elph_hydrogen_sulfide_hybrid}. 
Given the ubiquity of EPC in condensed matter, it is of primary importance to determine whether the much-used semi-local DFT only fails in the above-mentioned examples, or whether it is also insufficient in a wider context. 

The intrinsic computational expense of many-body methods makes EPC calculations prohibitive in many cases. Linear response methods~\cite{RevModPhys.73.515}, which are computationally advantageous, are not yet available in the context of many-body perturbation theory. Instead, the existing EPC studies beyond semi-local DFT rely on the frozen-phonon approach, which is computationally expensive.   
In a first attempt to reduce the computational cost of these calculations, Faber and co-workers~\cite{gw_gradients_blase} have explored approximations to the $GW$ self-energy ionic gradients needed for a perturbative calculation of EPC effects, demonstrating that the constant screening approach (using the screened Coulomb potential of the static lattice for the distorted structures) is a viable alternative to obtain relatively accurate results in carbon-based materials.

In this paper, I present an accurate and computationally inexpensive method to include many-body effects in the calculation of the temperature dependence of band structures, one of the signatures of EPC. The prediction of accurate band gaps and their temperature dependence is central in fields ranging from photovoltaics~\cite{shockley_queisser_limit} to topological insulators~\cite{elph_topological_prl,elph_topological_prb,elph_topological_jhi}. The proposed method relies on the exploration of the vibrational phase space along the recently introduced thermal lines~\cite{thermal_lines}, and accurate results are obtained with a \textit{single} many-body calculation at each temperature of interest. 

As a first application of the method, the $G_0W_0$ approximation is used to calculate the temperature dependence of the direct gaps of diamond (C), silicon (Si), lithium fluoride (LiF), magnesium oxide (MgO), and rutile titanium dioxide (TiO$_2$). The many-body contributions to the strength of EPC beyond semi-local DFT are found to be about $50$\% in C, Si, and TiO$_2$, but below $5$\% in LiF and MgO. These results reveal a complex interplay between electron correlation and electron-phonon coupling.


The vibrations in a solid formed by $N$ atoms can be described by the $3(N-1)$ harmonic normal coordinates $\{q_{\nu\mathbf{k}}\}$, where $\mathbf{k}$ represents a vibrational Brillouin zone (BZ) point, and $\nu$ a branch index. General configurations of the atoms in the solid are described by the vector $\mathbf{q}$, a $3(N-1)$-dimensional vector containing the amplitude of each normal coordinate. The electronic band gap $E_{\mathrm{g}}(\mathbf{q})$ is a function of the atomic positions, and its adiabatic harmonic vibrational average at temperature $T$ is
\begin{equation}
\langle E_{\mathrm{g}}(T) \rangle = \frac{1}{\mathcal{Z}}\sum_{\mathbf{M}}\langle\phi_{\mathbf{M}}(\mathbf{q})|E_{\mathrm{g}}(\mathbf{q})|\phi_{\mathbf{M}}(\mathbf{q})\rangle e^{-\frac{E_{\mathbf{M}}}{k_{\mathrm{B}}T}}, \label{eq:average}
\end{equation}
where $|\phi_{\mathbf{M}}\rangle=\prod_{\nu,\mathbf{k}}|\phi_{M_{\nu\mathbf{k}}}(q_{\nu\mathbf{k}})\rangle$ is a vibrational state of energy $E_{\mathbf{M}}$, $\mathbf{M}$ contains all the relevant quantum numbers, $\mathcal{Z}=\sum_{\mathbf{M}} e^{-E_{\mathbf{M}}/k_{\mathrm{B}}T}$ is the partition function, and $k_{\mathrm{B}}$ is the Boltzmann constant. 
The vibrational average in Eq.~(\ref{eq:average}) has been calculated using molecular dynamics~\cite{galli_water_gap_aimd} and path integral molecular dynamics~\cite{PhysRevB.73.245202,ceperley_h_elph_coupling}, Monte Carlo integration~\cite{giustino_nat_comm,helium}, and perturbative expansions over small atomic displacements, known as Allen-Heine-Cardona theory or the quadratic approximation~\cite{PhysRevLett.105.265501,gonze_off_diagonal,PhysRevB.87.144302,elph_Si_nano,gonze_marini_elph,monserrat_elph_diamond_silicon,gonze_gw_elph}. All these methods involve the evaluation of the band gap at multiple atomic configurations $\mathbf{q}$, of the order of $100$ to $1000$, and their use in conjunction with computationally intensive many-body techniques is therefore very limited.

Motivated by the mean value theorem for integrals, a simpler method to evaluate Eq.~(\ref{eq:average}) can be found following Ref.~\onlinecite{thermal_lines} to define a set of $2^{3(N-1)}$ thermal lines $\mathcal{T}_{\mathbf{S}}(T)$ in configuration space, parametrized by the temperature $T$, for which atomic positions are given by:
\begin{equation}
q_{\nu\mathbf{k}}(T)=\pm\left(\frac{1}{2\omega_{\nu\mathbf{k}}}[1+2 n_{\mathrm{B}}(\omega_{\nu\mathbf{k}},T)]\right)^{1/2}, \label{eq:tl}
\end{equation}
where $n_{\mathrm{B}}(\omega,T)=(e^{\omega/k_{\mathrm{B}}T}-1)^{-1}$ is a Bose-Einstein factor. Thermal lines $\mathcal{T}_{\mathbf{S}}$ are labelled by the $3(N-1)$-dimensional vector $\mathbf{S}$ that contains the sign of each degree of freedom in Eq.~(\ref{eq:tl}).
Thermal lines have the property that the value of a physical quantity at position $T$ is approximately equal to the value of the vibrational average of that quantity at temperature $T$. For example, the value of $\langle E_{\mathrm{g}}(T)\rangle$ is approximately equal to the value of $E_{\mathrm{g}}[\mathcal{T}_{\mathbf{S}}(T)]$ for an atomic configuration at position $T$ on any thermal line $\mathcal{T}_{\mathbf{S}}$. The value of a property at point $T=0$ is approximately equal to the quantum average of that property, and points $T=0$ are referred to as quantum points. For more details about thermal lines, see Ref.~\onlinecite{thermal_lines}.

Thermal lines can be used to efficiently explore the vibrational phase space to calculate quantum and thermal averages~\cite{thermal_lines}. 
Here, a \textit{mean thermal line} is defined such that the corresponding atomic configuration has a value of the band gap equal to the vibrational average, and thus a single many-body calculation on the mean thermal line is sufficient to obtain accurate vibrational averages. 


The calculations described in this paper have been performed using both semi-local DFT and the many-body $G_0W_0$ approximation. The DFT calculations have been performed using the plane-wave {\sc quantum espresso} package~\cite{quantum_espresso}. 
The DFT results provide the starting Kohn-Sham states used in the many-body calculations performed with the plane-wave {\sc Yambo} code~\cite{yambo}. 
The numerical parameters of the calculations are described in the Supplemental Material~\cite{supp_tl_gw_arxiv}, and they lead to static lattice direct gaps as shown in Table~\ref{tab:latt_param}, in agreement with literature values~\cite{gw_si_c_gas_ge_sic,Pasquarello_GW_semiconductors}. 
The lattice parameters of the different structures are also shown in Table~\ref{tab:latt_param}. The vibrational eigenmodes have been obtained using the finite displacement method~\cite{phonon_finite_displacement}.

\begin{table}
\setlength{\tabcolsep}{10pt} 
\caption{Lattice parameters and static lattice $G_0W_0$ direct band gaps of the materials studied in this work.} 
\label{tab:latt_param}
\begin{tabular}{cccc}
\hline
\hline
System  & $a$ (\AA) &  $c$ (\AA) & $E_{\mathrm{g}}$ (eV) \\
\hline
C         & $3.532$ & --      &  $7.49$ \\
Si        & $5.394$ & --      &  $3.32$ \\
LiF       & $4.068$ & --      &  $12.98$ \\
MgO       & $4.251$ & --      & $6.61$ \\
TiO$_2$   & $4.587$ & $2.949$ & $3.26$ \\
\hline
\hline
\end{tabular}
\end{table}



\begin{figure}
\centering
\includegraphics[scale=0.40]{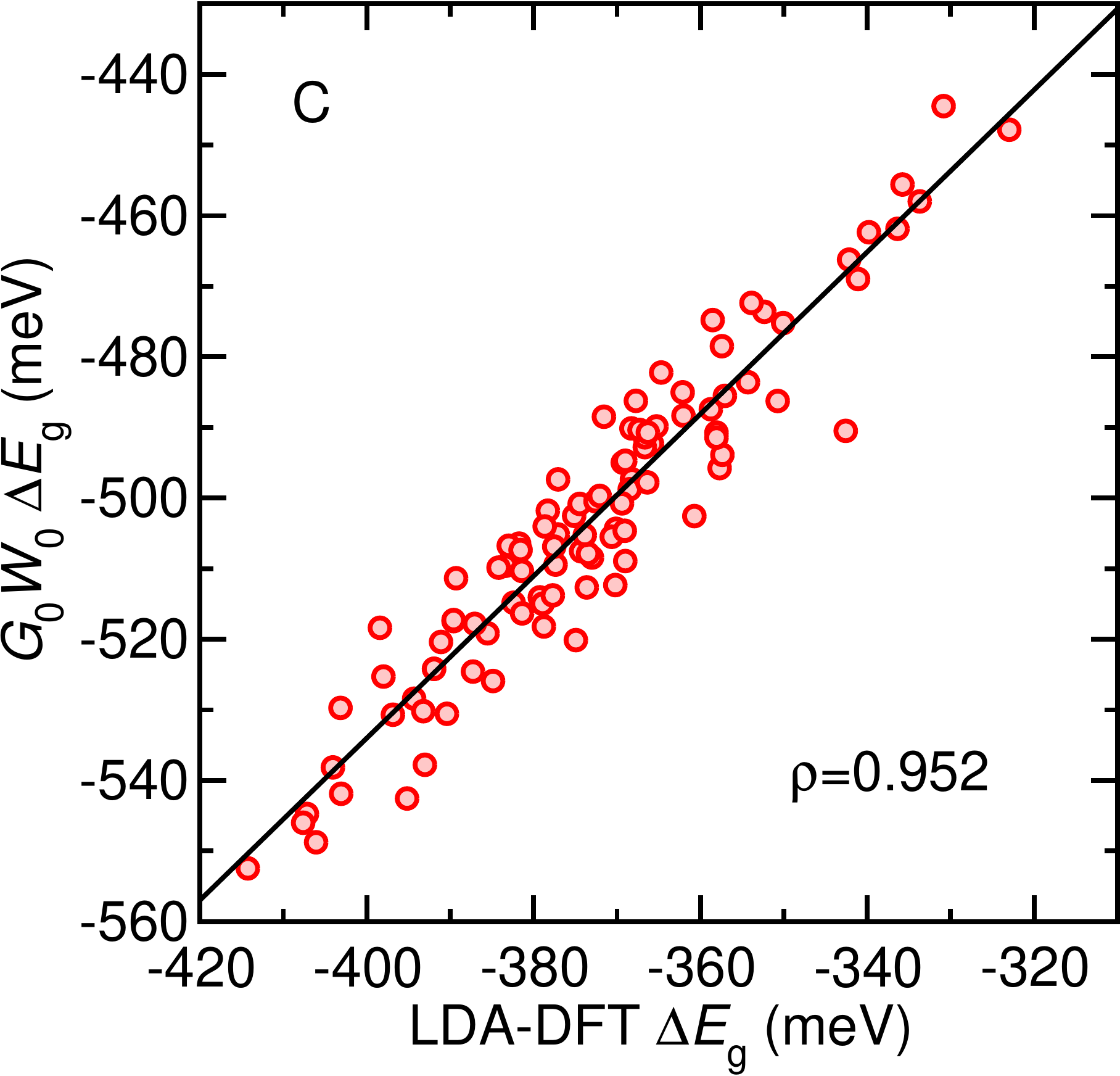}
\caption{$G_0W_0$ band gap ZP correction as a function of the LDA-DFT band gap ZP correction for a $54$-atom simulation cell of C. The correlation coefficient is $\rho=0.952$ and the RMS of the residuals of a straight line fit is $6.8$~meV.}
\label{fig:correlation}
\end{figure}

The methodology is initially described using C as a test system, as previous calculations using alternative methods exist for comparison, and the results for the other materials are presented later. A set of $100$ thermal lines is randomly chosen for C, and their quantum points ($T=0$) are used in the following. For each of the $100$ corresponding atomic configurations, the optical gap $E_{\mathrm{g}}[\mathcal{T}_{\mathbf{S}}(0)]$ is calculated using the local density approximation of density functional theory (LDA-DFT) and the $G_0W_0$ approximation. Figure~\ref{fig:correlation} shows the band gap zero-point (ZP) correction $\Delta E_{\mathrm{g}}=E_{\mathrm{g}}[\mathcal{T}_{\mathbf{S}}(0)]-E_{\mathrm{g}}(\mathbf{0})$ for each of the $100$ configurations, comparing the $G_0W_0$ ZP correction (vertical axis) against the LDA-DFT ZP correction (horizontal axis). The central feature of Fig.~\ref{fig:correlation} is the strong linear correlation between the band gap corrections evaluated with either method. Most studies assume that many-body effects on electron-phonon coupling are negligible, and such scenario, which is shown to hold for LiF and MgO below, would lead to a linear correlation with a slope of one and zero offset. The present work shows that, for some materials, including C in Fig.~\ref{fig:correlation}, the above scenario breaks down and many-body effects make an important contribution to EPC. 
The results depicted in Fig.~\ref{fig:correlation} correspond to a $54$-atom simulation cell, but similar correlations are observed at other cell sizes. 

The linear correlation between the $G_0W_0$ and the LDA-DFT corrections suggests a simple scheme: (i) choose an appropriate thermal line at the DFT level, and (ii) use this single thermal line to explore configuration space using many-body calculations. Thus, $n$ thermal lines are sampled at the DFT level at small computational cost, and then the mean thermal line $\mathcal{T}_{\overline{\mathbf{S}}}$ is selected as:
\begin{equation}
\overline{\mathbf{S}}_n=\argmin_{\mathbf{S}}|E_{\mathrm{g}}[\mathcal{T}_{\mathbf{S}}(0)]-\langle E_{\mathrm{g}}(0)\rangle|. \label{eq:minimum}
\end{equation}
The correlation in Fig.~\ref{fig:correlation} ensures that $\overline{\mathbf{S}}_n^{\mathrm{DFT}}\simeq\overline{\mathbf{S}}_n^{G_0W_0}$. This procedure can be iterated to define $i$th order mean thermal lines, which could be used to increase the accuracy at small additional computational cost, as demonstrated for TiO$_2$ below. In Eq.~(\ref{eq:minimum}) the quantum point $T=0$ is used to select the mean thermal line, but similar results are obtained at other temperatures, as thermal lines at different temperatures are correlated~\cite{thermal_lines}. As semi-local DFT calculations are computationally inexpensive compared to many-body calculations, a thermal line can be defined at each temperature of interest with a new semi-local DFT sampling. This approach has been used below for TiO$_2$.

\begin{figure}
\centering
\includegraphics[scale=0.40]{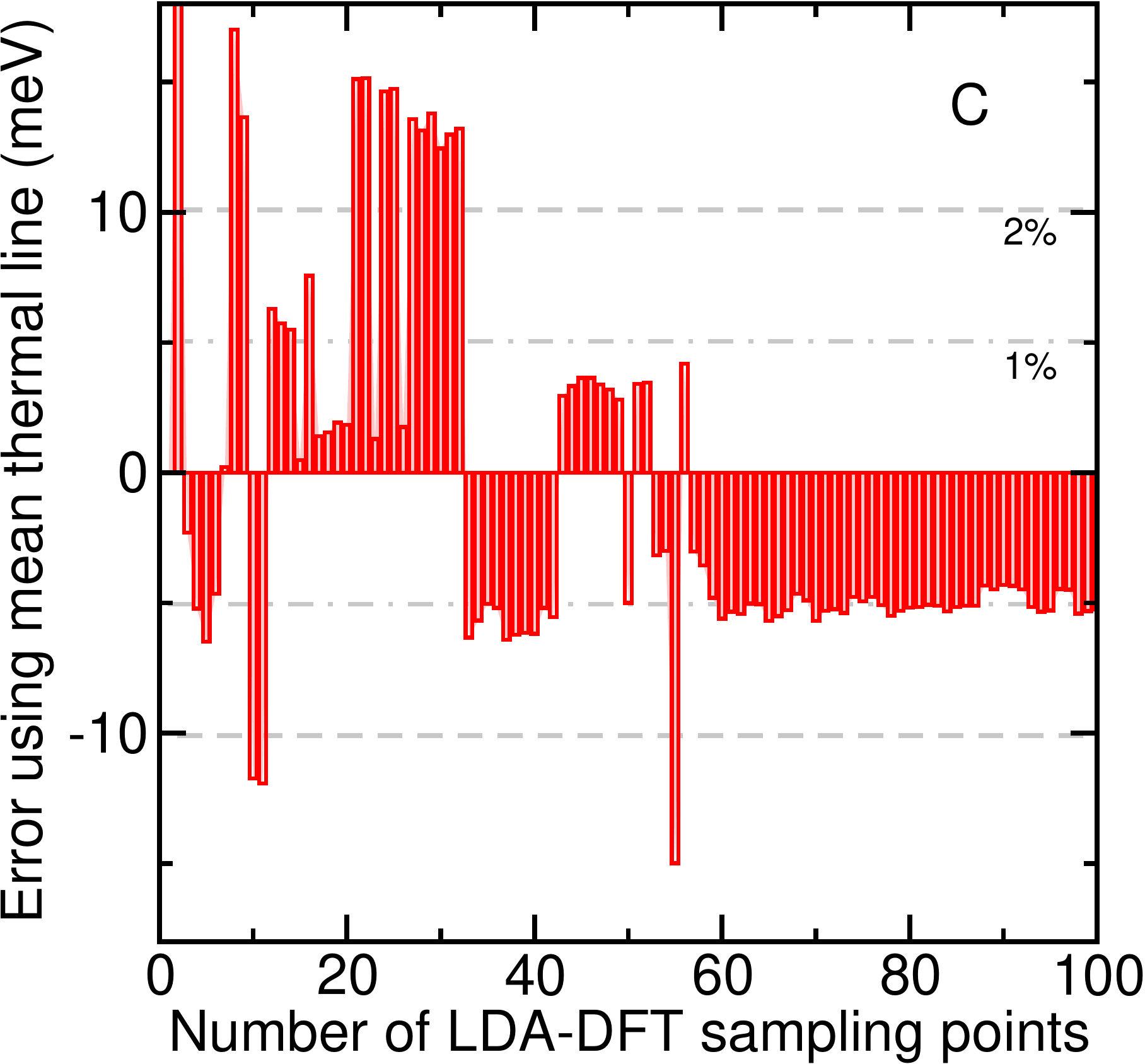}
\caption{Difference between the $G_0W_0$ ZP correction calculated by using the mean thermal line $E_{\mathrm{g}}[\mathcal{T}_{\overline{\mathbf{S}}_n}(0)]$ and by averaging over a set of thermal lines $\langle E_{\mathrm{g}}(0)\rangle_n$. The difference is plotted as a function of the number of thermal lines $n$ used at the LDA-DFT level to select the mean thermal line according to Eq.~(\ref{eq:minimum}). The grey dashed and dotted-dashed lines indicate $2$\% and $1$\% errors respectively.}
\label{fig:error}
\end{figure}

To test the methodology, the $G_0W_0$ ZP correction to the optical gap of C is evaluated using the data for the $54$-atom simulation cells used in Fig.~\ref{fig:correlation}. First, the $G_0W_0$ ZP correction is calculated by averaging over $n$ thermal lines, $\langle E_{\mathrm{g}}(0)\rangle_n=\frac{1}{n}\sum_{i=1}^nE_{\mathrm{g}}[\mathcal{T}_{\mathbf{S}_i}(0)]$, as in Ref.~\onlinecite{thermal_lines}. Second, the same ZP correction is evaluated by selecting the mean thermal line from $n$ LDA-DFT data points according to Eq.~(\ref{eq:minimum}), and evaluating the $G_0W_0$ ZP correction only for the configuration corresponding to the mean thermal line, $E_{\mathrm{g}}[\mathcal{T}_{\overline{\mathbf{S}}_n}(0)]$. The results are summarized in Fig.~\ref{fig:error}. For a given number $n$ of thermal lines (horizontal axis), Fig.~\ref{fig:error} shows $E_{\mathrm{g}}[\mathcal{T}_{\overline{\mathbf{S}}_n}(0)]-\langle E_{\mathrm{g}}(0)\rangle_n$, the difference between the $G_0W_0$ ZP correction calculated using only the mean thermal line and that calculated by averaging over thermal lines. Errors below $3$\% are achieved using as few as $10$ thermal lines to determine the mean thermal line. The residual error of $1\%$ arises from the finite spread of the points in Fig.~\ref{fig:correlation}, and is similar to the size of the RMS. 

For the full data set ($n=100$), the $G_0W_0$ ZP band gap correction is $-508$~meV using the mean thermal line, and the average over $100$ thermal lines is $-503$~meV ($1$\% error). 
To put this error in context, in C there is a $50$~meV uncertainty ($12$\% error) associated with the choice of pseudopotential~\cite{gonze_marini_elph}, and up to a $40$~meV uncertainty ($9$\% error) associated with the use of a quadratic or a Monte Carlo method to evaluate Eq.~(\ref{eq:average})~\cite{thermal_lines}. Therefore, using the mean thermal line $\mathcal{T}_{\overline{\mathbf{S}}}$ appears to be a promising approach to routinely and accurately calculate the temperature dependence of band structures including many-body effects.

\begin{figure}
\centering
\includegraphics[scale=0.40]{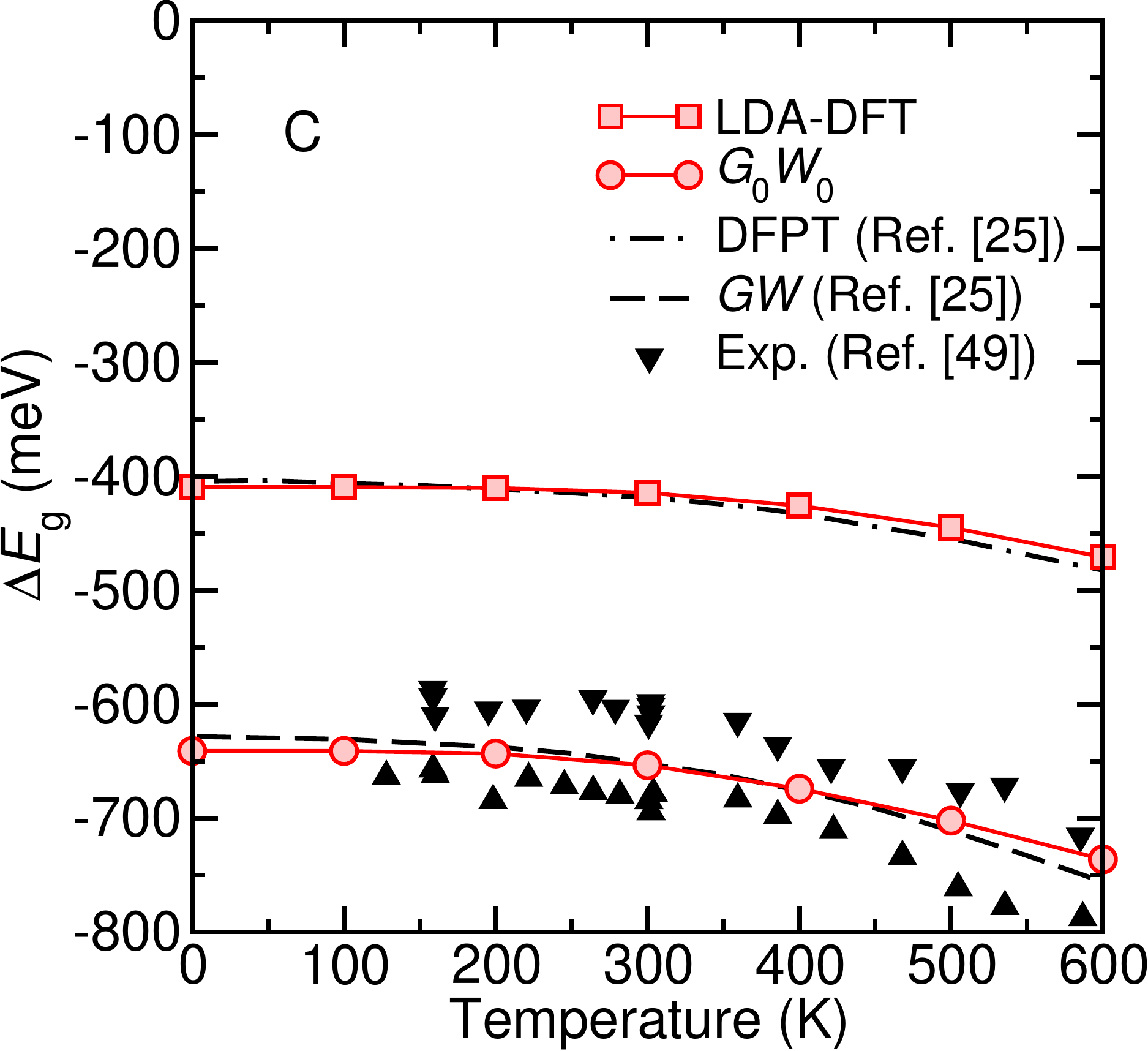}
\caption{ZP correction and temperature dependence of the direct gap of C. The red squares correspond to the LDA-DFT calculation, and the red circles to the $G_0W_0$ calculation. 
The dotted-dashed and dashed black lines are theoretical results from Ref.~\onlinecite{gonze_gw_elph}, and the black triangles are experimental results from Ref.~\onlinecite{diamond_tdependence_experiment}. For the comparison, the static lattice value of the band gap has been taken from Ref.~\onlinecite{gonze_gw_elph} to be $7.732$~eV. This differs from the value reported in Tab.~\ref{tab:latt_param} because the self-consistent $GW$ method was used in Ref.~\onlinecite{gonze_gw_elph}, compared to the one-shot $G_0W_0$ in the present work.
}
\label{fig:tdep}
\end{figure}

The final results for C are presented in Fig.~\ref{fig:tdep} for a larger simulation cell containing $128$ atoms. 
The calculations have been performed by randomly sampling $100$ thermal lines using LDA-DFT, then selecting the mean thermal line $\mathcal{T}_{\overline{\mathbf{S}}}$ according to Eq.~(\ref{eq:minimum}), and performing a single many-body $G_0W_0$ calculation at each temperature of interest along the mean thermal line. The results in Fig.~\ref{fig:tdep} are in good agreement with previous theoretical calculations~\cite{gonze_gw_elph} (black lines), and with available experimental data (black triangles)~\cite{diamond_tdependence_experiment}. 

The present approach requires a total of $7$ $G_0W_0$ calculations for $7$ different temperatures. In the standard quadratic approximation using the frozen-phonon method~\cite{gonze_gw_elph}, $62$ $GW$ calculations are required~\cite{antonius_communication}. For a $4\times4\times4$ BZ sampling grid, an equivalent supercell contains $128$ atoms, and should naively require $3\times(128-1)=381$ sampling points. However, exploiting the symmetries of the diamond structure, this number can be reduced to $62$ sampling points. The method presented in this work requires an order of magnitude fewer calculations than the quadratic method. The reduction in computational cost is smaller, as in the quadratic method not all points in the BZ grid require the same supercell size. A point $(m_1/n_1,m_2/n_2,m_3/n_3)$ (in fractional coordinates) can be accessed with supercells of size $n_1n_2n_3$ using the standard supercell approach, or smaller supercells of size equal to the least common multiple of $n_1$, $n_2$, and $n_3$, using non-diagonal supercells~\cite{non_diagonal}.

\begin{figure}
\centering
\includegraphics[scale=0.40]{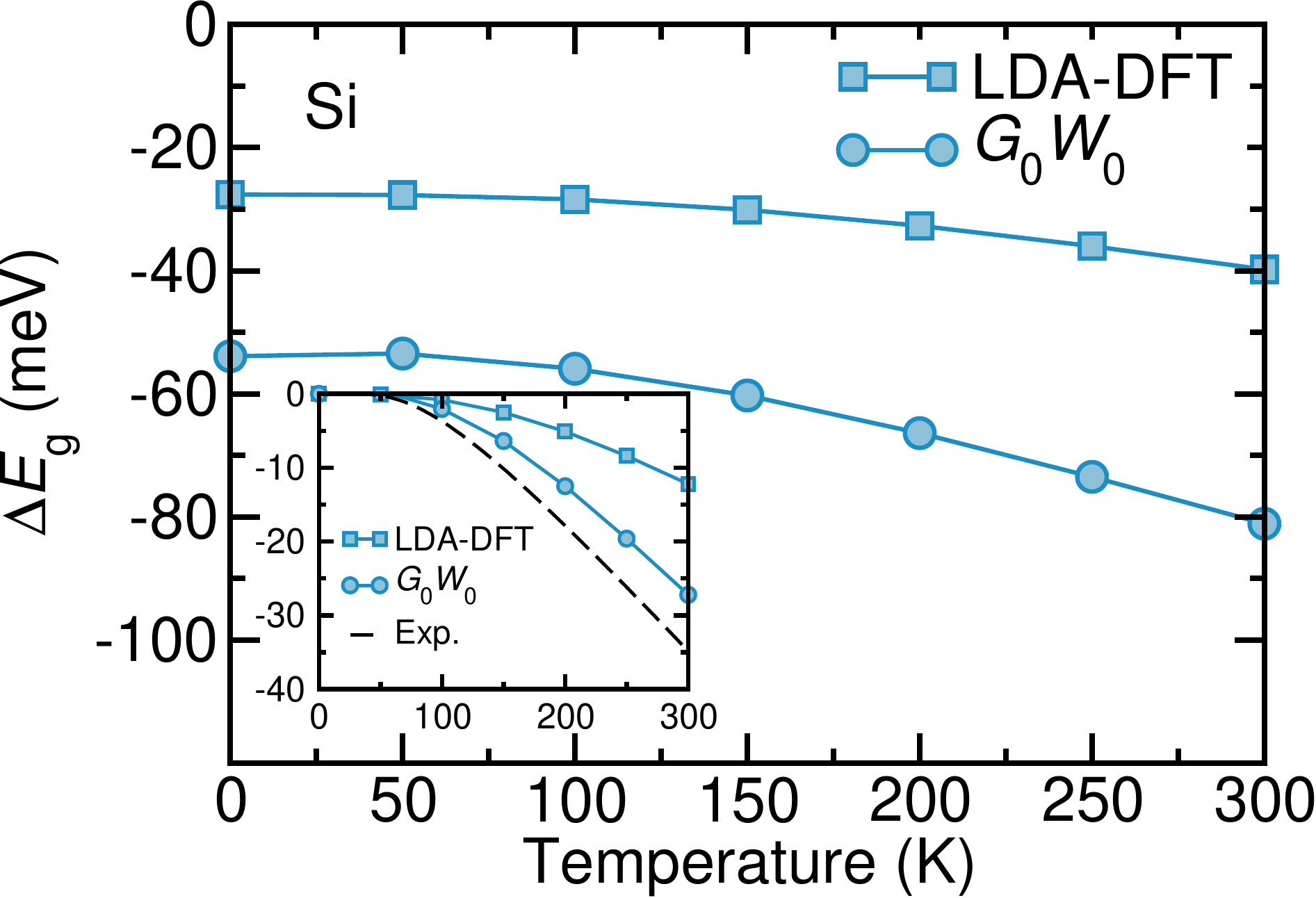}
\includegraphics[scale=0.40]{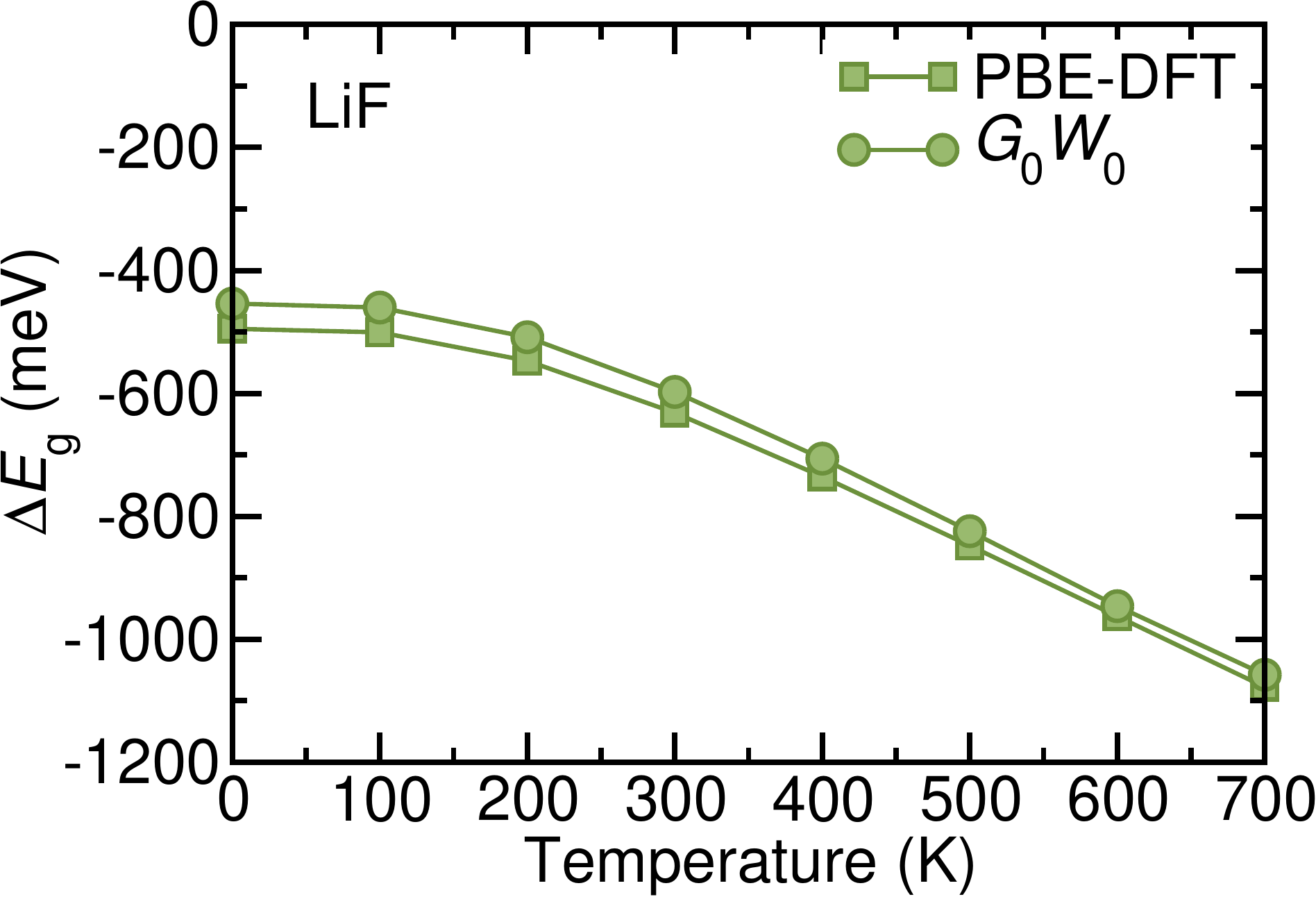}
\includegraphics[scale=0.40]{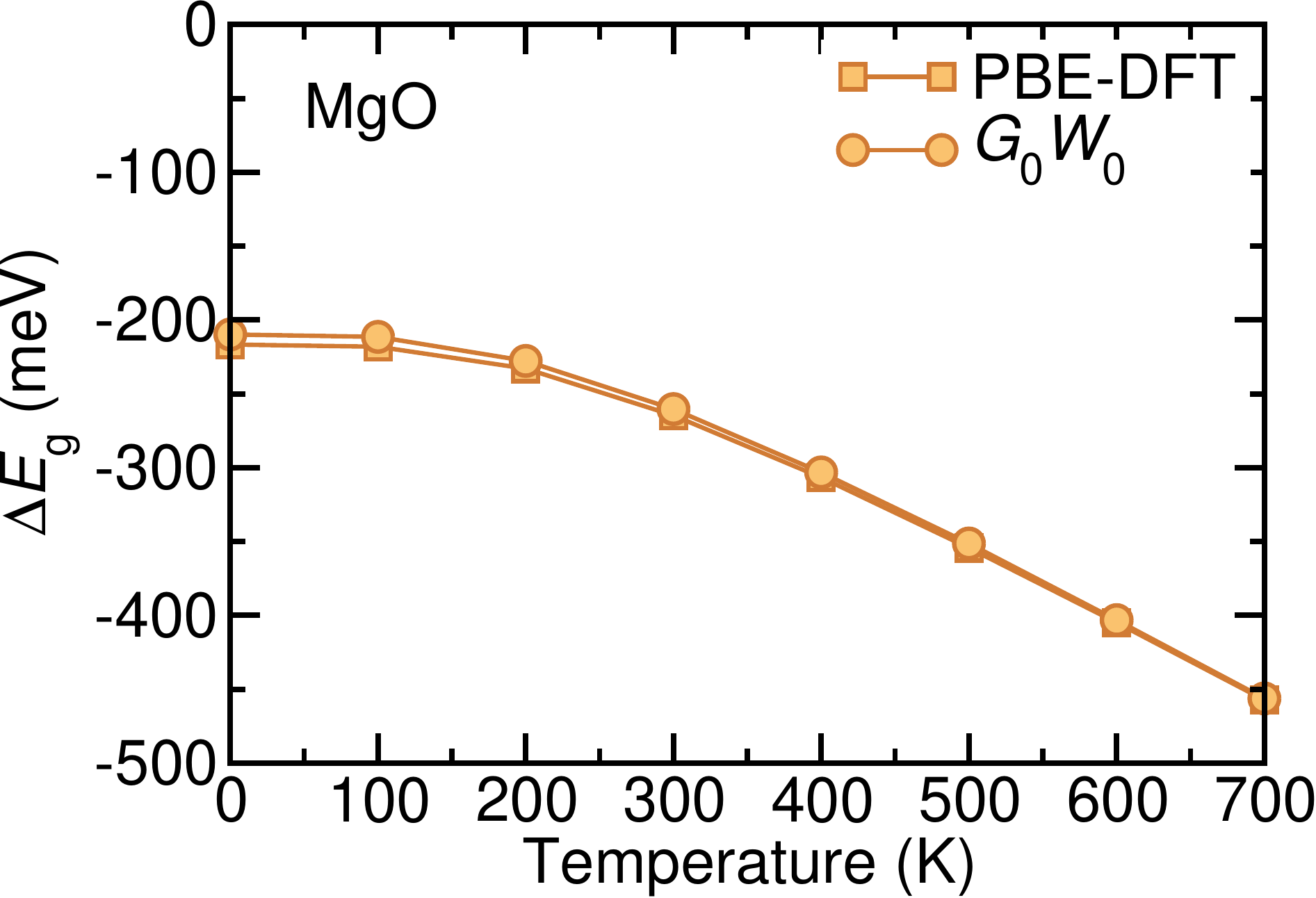}
\includegraphics[scale=0.40]{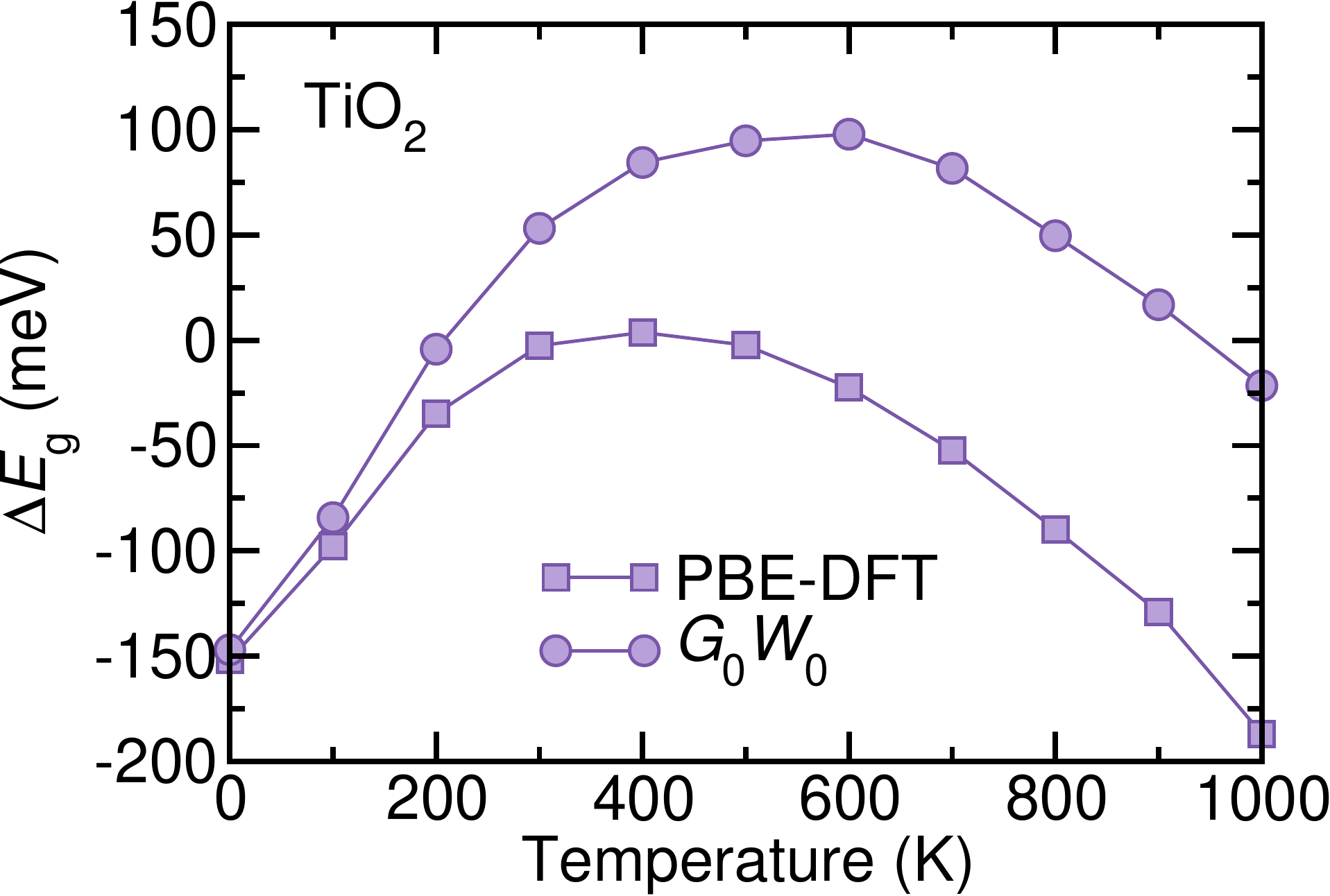}
\caption{ZP correction and temperature dependence of the direct gaps of Si, LiF, MgO, and TiO$_2$. The squares correspond to the DFT calculations, and the circles to the $G_0W_0$ calculations. The inset in the Si graph shows a comparison to experimental data (black dashed line) from Ref.~\onlinecite{silicon_exp_tdep}.} 
\label{fig:si_mgo_tio2}
\end{figure}

Linear correlations between the DFT and $G_0W_0$ ZP corrections to the gap are also observed in Si, LiF, MgO, and TiO$_2$ (see Supplemental Material~\cite{supp_tl_gw_arxiv}). The method proposed here relies on this correlation, for which there is no general proof. However, it has also been reported in molecular crystals between semi-local DFT and hybrid functional DFT~\cite{molec_crystals_elph}, suggesting that mean thermal lines for the inclusion of many-body effects will be applicable in a wide variety of materials.

In Fig.~\ref{fig:si_mgo_tio2}, the ZP correction and temperature dependence of the direct gaps of these four materials are reported. The calculations have been performed using the mean thermal line, selected from $100$ DFT thermal lines, and with simulation cells containing $128$ atoms for Si, LiF, and MgO, and $108$ atoms for TiO$_2$. For TiO$_2$, the DFT ZP corrections have a wide range of values, and the accuracy of the results is significantly improved by using higher order mean thermal lines. 

For Si, the inset in Fig.~\ref{fig:si_mgo_tio2} shows a comparison with the experimental temperature dependence~\cite{silicon_exp_tdep}, where the zero-temperature band gap has been taken as a reference. As in diamond, the agreement with experiment improves significantly with the inclusion of electron correlation, and the remaining difference could be attributed to finite-size effects~\cite{jcp_ponce_convergence}. For LiF and MgO, the inclusion of correlation effects decreases the band gap correction slightly. In TiO$_2$, the temperature dependence of the band gap is nonmonotonic, with low energy phonons excited at low temperatures opening the gap, and harder, higher energy phonons closing the gap, with a cross-over between the two regimes at about $400$~K using DFT, and $600$~K using $G_0W_0$. This difference provides a good test of the theory by experiment. In TiO$_2$, even though the ZP correction to the gap is very similar between DFT and $G_0W_0$, the distinct temperature dependences demonstrate that correlation significantly affects electron-phonon coupling in this material. Finally, it should be noted that excitonic effects~\cite{marini_finite_temp_excitons} are important in the optical properties of some materials, and that in polar compounds, $GW$ band gaps can be modified by long-range electric fields~\cite{polarisation_gw_gap_renormalisation} and electron-phonon coupling strength is also affected~\cite{verdi_polar_elph,sjakste_polar_elph}. The inclusion of these effects is beyond the scope of the present work. Lattice thermal expansion also modifies band gaps, as discussed in the Supplemental Material~\cite{supp_tl_gw_arxiv}. 



In conclusion, I have introduced an accurate and computationally inexpensive method to evaluate the effects of electron-phonon coupling on band structures using many-body techniques along thermal lines. 
Although the results presented here have been obtained using the one shot $G_0W_0$ approximation, the same methodology should be applicable to other many-body techniques, such as self-consistent $GW$, or to wave function methods such as quantum Monte Carlo. 
The temperature dependence of the band gaps of diamond, silicon, lithium fluoride, magnesium oxide, and titanium dioxide, has been calculated using the $G_0W_0$ approximation. Many-body contributions range from $5$ to $50$\% in these materials, suggesting a complex interplay between electron-electron and electron-phonon coupling. 

\acknowledgements

The author thanks Gabriel Antonius for providing details about the calculations in Ref.~\onlinecite{gonze_gw_elph}, Feliciano Giustino for useful conversations, Mireia Crispin-Ortuzar for helpful comments on the manuscript, and Robinson College, Cambridge, and the Cambridge Philosophical Society for a Henslow Research Fellowship.

\bibliography{/Users/bartomeumonserrat/Documents/research/papers/references/anharmonic}


\widetext
\clearpage

\begin{center}
\textbf{\large Supplemental Material for ``Correlation effects on electron-phonon coupling in semiconductors: many-body theory along thermal lines''}
\end{center}

\setcounter{equation}{0}
\setcounter{figure}{0}
\setcounter{table}{0}
\setcounter{page}{1}
\makeatletter
\renewcommand{\theequation}{S\arabic{equation}}
\renewcommand{\thefigure}{S\arabic{figure}}
\renewcommand{\bibnumfmt}[1]{[S#1]}
\renewcommand{\citenumfont}[1]{S#1}

\section{Computational details}

The calculations of the temperature dependence of the band gaps of C, Si, LiF, MgO, and TiO$_2$ have been performed using both semi-local DFT and the many-body $G_0W_0$ approximation.

The DFT calculations have been performed using the plane-wave {\sc quantum espresso} package~\cite{quantum_espresso}, with the local density approximation (LDA)~\cite{PhysRevLett.45.566,PhysRevB.23.5048} for C and Si, and the generalized gradient approximation of Perdew-Burke-Ernzerhof (PBE)~\cite{PhysRevLett.77.3865} for LiF, MgO, and TiO$_2$. Electronic Brillouin zone (BZ) grids of size $16\times16\times16$ centred at $\Gamma$ were used for the primitive $2$-atom cells of C, Si, LiF, and MgO, and a grid size of $10\times10\times12$ for the $6$-atom TiO$_2$ primitive cell. Commensurate grids have been used for the supercell calculations. Plane-wave energy cut-offs of $50$~Ha, $40$~Ha, $250$~Ha, $200$~Ha, and $200$~Ha have been chosen for C, Si, LiF, MgO, and TiO$_2$, respectively. 
Scalar relativistic norm-conserving Troullier-Martins pseudopotentials~\cite{troullier_martins_pseudo} from the {\sc quantum espresso} library are used for all elements, with the inclusion of semi-core states for Ti. For comparison, a calculation using a $54$-atom simulation cell yields a LDA-DFT zero-point (ZP) correction to the direct gap of diamond of $-373$~meV, in good agreement with the $-381$~meV correction obtained for the same structure using Vanderbilt ultrasoft pseudopotentials~\cite{PhysRevB.41.7892} and the {\sc castep} code~\cite{castep} (differences of up to $50$~meV due to the choice of pseudopotential have been reported for diamond~\cite{gonze_marini_elph}).

The DFT results provide the starting Kohn-Sham states used in the many-body calculations performed with the plane-wave {\sc Yambo} code~\cite{yambo}. The $G_0W_0$ calculations have been performed within the plasmon-pole approximation~\cite{plasmon_pole_approx}, with cutoffs of $7$~Ha (C, MgO, and TiO$_2$), $4$~Ha (Si), and $9$~Ha (LiF) for the dielectric function, and including a number of bands appropriate to obtain converged results at each system size. The resulting static lattice direct gaps of $7.49$~eV, 
$3.32$~eV, 
$12.98$~eV, $6.61$~eV, and $3.26$~eV for C, Si, LiF, MgO, and TiO$_2$, respectively, are in agreement with literature values~\cite{gw_si_c_gas_ge_sic,Pasquarello_GW_semiconductors}. Less stringent convergence criteria can be used for the evaluation of the energy differences required for the calculation of band gap corrections.

The structures and vibrational eigenmodes used in the present work are the same as those in Ref.~\cite{thermal_lines} for C and Si, with lattice parameters of $3.532$~\AA~and $5.394$~\AA~respectively. For LiF, the relaxed structure within PBE has a lattice parameter of $4.068$~\AA\@, for MgO of $4.251$~\AA\@, and for TiO$_2$, a structure with lattice parameters $a=4.587$~\AA~and $c=2.949$~\AA~was used instead of the PBE relaxed structure, as the latter leads to imaginary phonons. The vibrational eigenmodes have been obtained using the finite displacement method~\cite{phonon_finite_displacement}.

\section{Relation between DFT and $G_0W_0$ for {Si, LiF, MgO, and TiO}$_2$}

\begin{figure}
\centering
\includegraphics[scale=0.4]{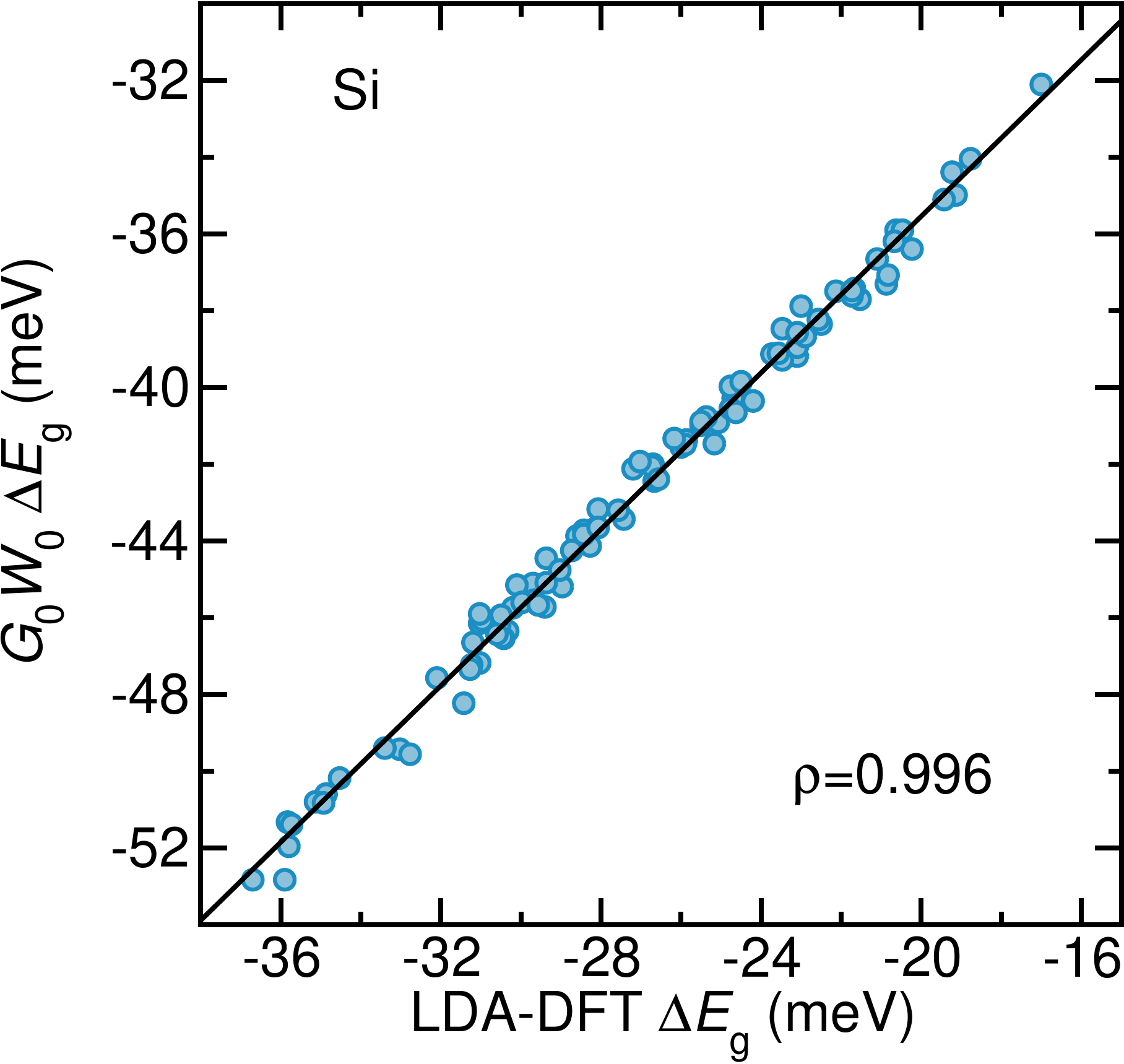}
\includegraphics[scale=0.4]{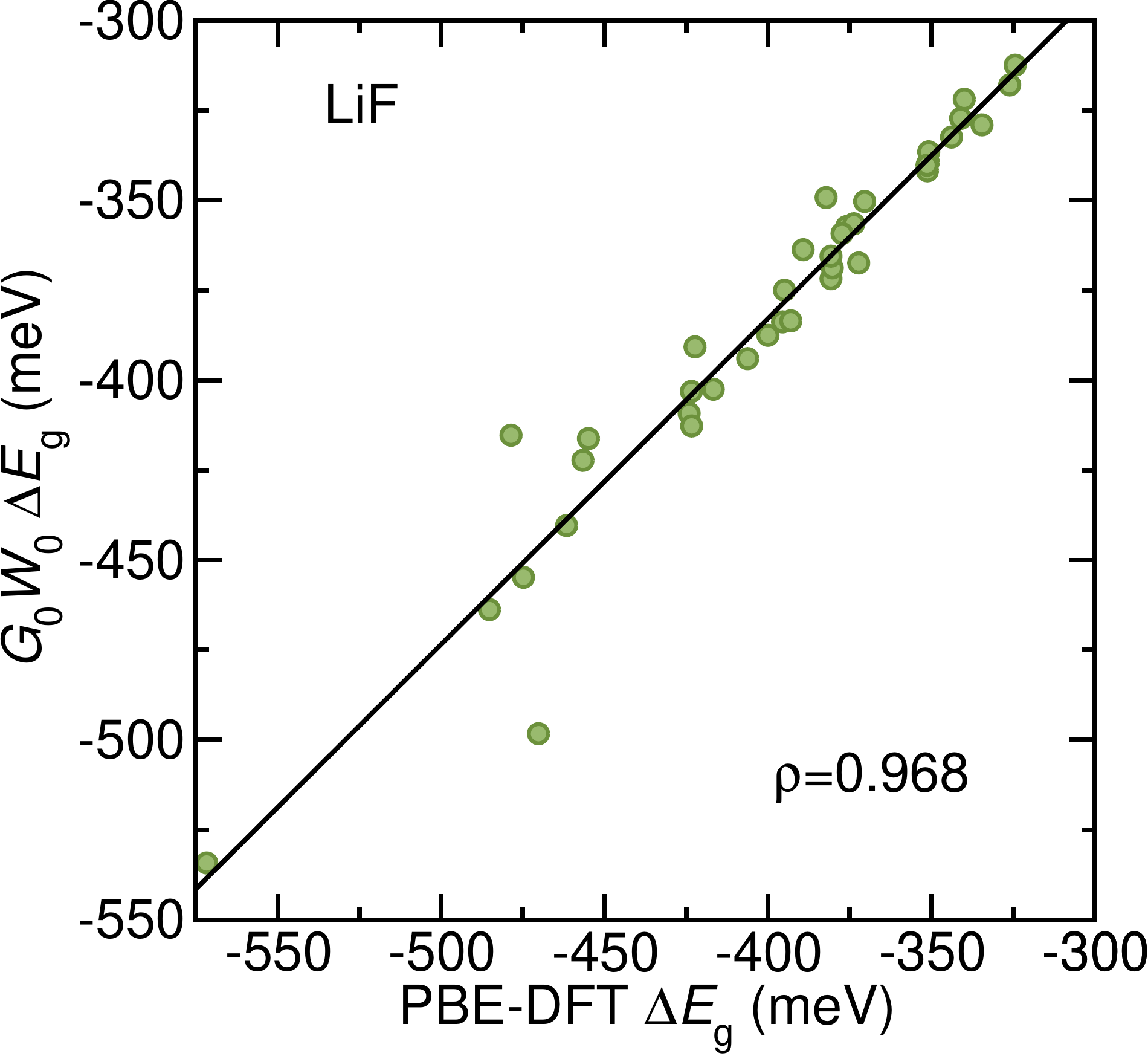}
\includegraphics[scale=0.4]{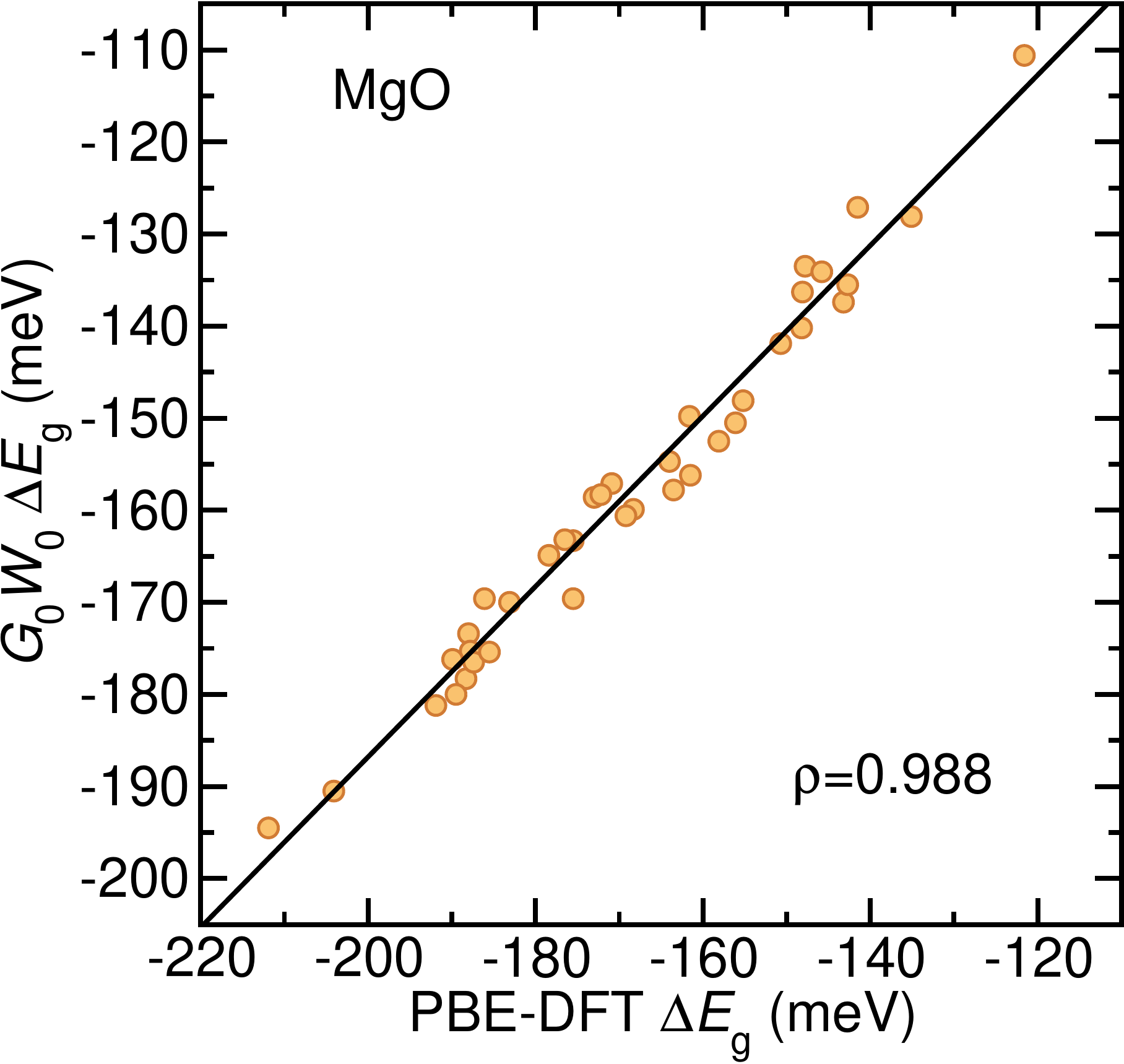}
\includegraphics[scale=0.4]{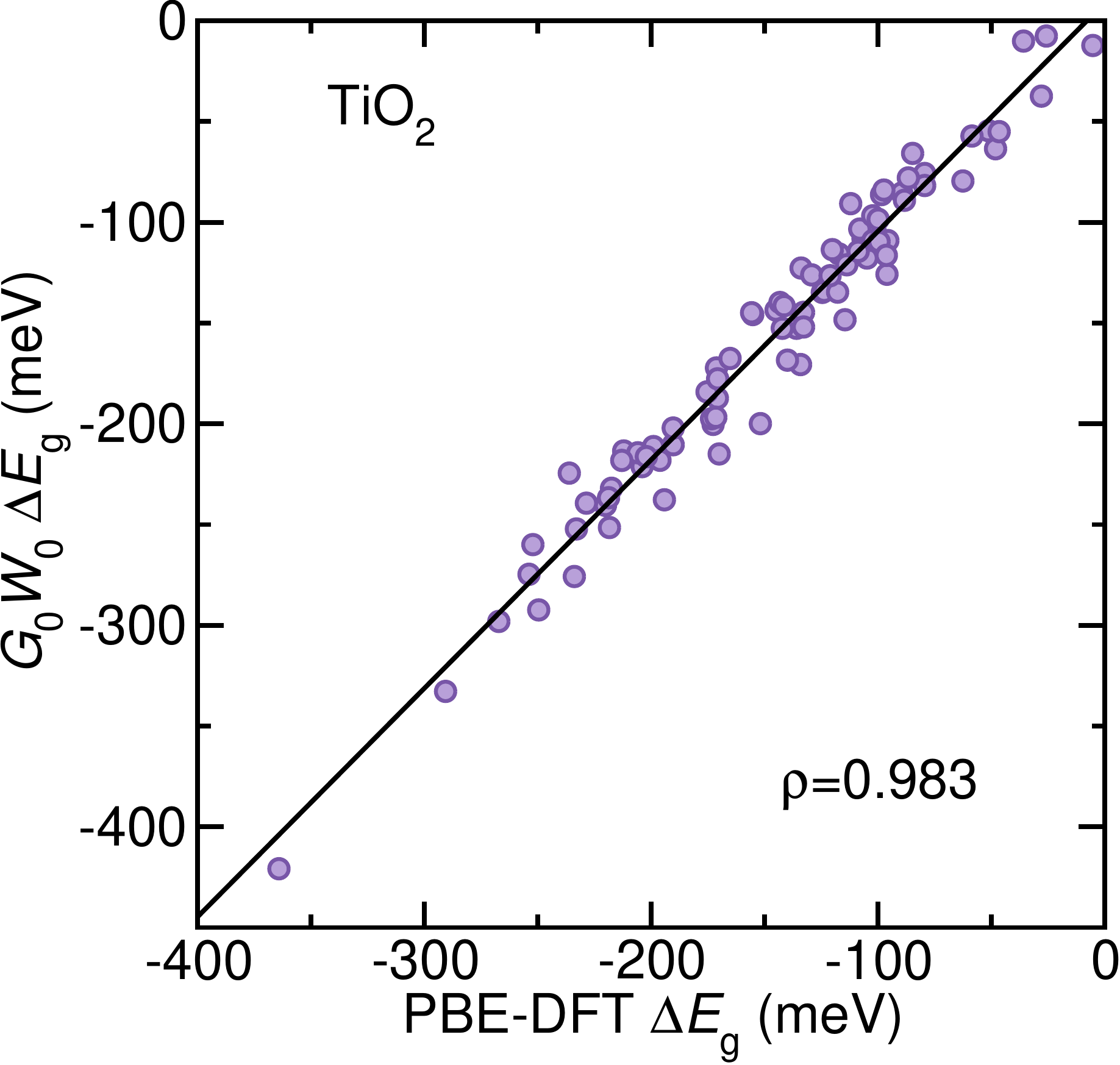}

\caption{$G_0W_0$ band gap ZP correction as a function of the DFT band gap ZP correction. The results for Si, LiF, and MgO were obtained using a $54$-atom simulation cell. The results for TiO$_2$ were obtained with a simulation cell containing $108$ atoms.}
\label{fig:correlation}
\end{figure}

Figure~\ref{fig:correlation} shows the $G_0W_0$ band gap ZP correction as a function of the DFT band gap ZP correction for Si, LiF, MgO, and TiO$_2$. A strong correlation is observed with correlation coefficients above $0.96$ in all cases.


\section{Thermal expansion}

Thermal expansion affects the size of the direct gap of materials. It is a separate effect to the change induced by electron-phonon coupling, and the conclusions of the main manuscript regarding the validity of semi-local DFT compared to many-body perturbation theory are independent of it. Nonetheless, it is interesting to estimate its size.

In diamond, the gap decreases by $16$~meV when temperature increases from $0$~K to $1000$~K~\cite{PhysRevB.87.144302}. This change is over an order of magnitude smaller than the change caused by electron-phonon coupling in this temperature range.

For the rest of materials, the effects of thermal expansion on the gap are estimated by considering experimental volumes as a function of temperature. Silicon has a nonmonotonic thermal expansion behaviour, exhibiting thermal contraction at temperatures below about $110$~K~\cite{silicon_thermal_contraction}, and at $300$~K the lattice constant has increased by about $0.002$~\AA~from zero temperature~\cite{silicon_a}. As a result, thermal expansion closes the band gap in Si by $0.5$~meV at $300$~K, a negligible effect as in the case of diamond.

The gap closure induced by thermal expansion is significantly larger in the other materials. In LiF, the lattice parameter increases by $0.07$~\AA~at $700$~K~\cite{lif_thermal_expansion}, driving a gap clousure of $416$~meV. In MgO, the lattice parameter increases by $0.03$~\AA~at $700$~K~\cite{mgo_thermal_expansion}, accompanied by a gap closure of $192$~meV. In TiO$_2$, the $a$ lattice parameter increases by $0.02$~\AA, and the $c$ lattice parameter by $0.04$~\AA~at $700$~K~\cite{tio2_thermal_expansion}, while the band gap decreases by $63$~meV.

\end{document}